\documentclass[12pt,psamsfonts]{amsart}
\usepackage{amsmath}
\usepackage{amsthm}
\usepackage{amssymb}
\usepackage{amscd}
\usepackage{amsfonts}
\usepackage{amsbsy}
\usepackage{epsfig}

\newcommand{\Z}{\ensuremath{\mathbb{Z}}}

\begin{document}

\title[The gravitational quantization: The Titius--Bode Law]
{On the origin of the gravitational quantization: The Titius--Bode
Law}

\author[J. Gin\'{e}]
{Jaume Gin\'e}

\address{Departament de Matem\`atica, Universitat de Lleida,
Av. Jaume II, 69. 25001 Lleida, Spain}

\email{gine@eps.udl.es}

\thanks{The author is partially supported by a DGICYT
grant number BFM 2002-04236-C02-01 and by DURSI of Government of
Catalonia ``Distinci\'o de la Generalitat de Catalunya per a la
promoci\'o de la recerca universit\`aria".}


\date{}
\dedicatory{}

\maketitle

\begin{abstract}
Action at distance in Newtonian physics is replaced by finite
propagation speeds in classical post--Newtonian physics. As a
result, the differential equations of motion in Newtonian physics
are replaced by functional differential equations, where the delay
associated with the finite propagation speed is taken into
account. Newtonian equations of motion, with post--Newtonian
corrections, are often used to approximate the functional
differential equations. In \cite{G2} a simple atomic model based
on a functional differential equation which reproduces the
quantized Bohr atomic model was presented. The unique assumption
was that the electrodynamic interaction has finite propagation
speed. Are the finite propagation speeds also the origin of the
gravitational quantization? In this work a simple gravitational
model based on a functional differential equation gives an
explanation of the modified Titius--Bode law.
\end{abstract}

\section{Introduction}\label{s1}

In the last two centuries several attempts were made to express
and explain the distribution of the planetary orbits and other
relevant quantities using integer numbers. Titius (1772) and Bode
(1776) (see for instance \cite{LP, T}) proposed the law describing
the mean distances of planets from the Sun of the general form
\[
r_n= a+b \, c^n,
\]
where $r_n$ means distance characterized by an integer number $n$.
The constants $a$, $b$ and $c$ have no convincing physical
meaning, neither have the empirical correlations with definite
parameters for a given system. Therefore, this law has raised many
discussions. Nevertheless, it played a positive role not only in
predicting unknown planets, but also in stimulating many
researches to further work in this direction. In \cite{HT} and
\cite{RR1} there are some reviews about the attempts to test and
to explain the Titius-Bode law.

As a precursor of the modified Titius-Bode law, Gulak \cite{Gu1}
proposed that the orbital distances are given by $r_n= (n+1/2)r_0$
or $r_n=n r_0$, where $r_0$ is a characteristic of a given system.
Here $n$ needs not to increase by 1 in going from one planet or
satellite to another one. Gulak \cite{Gu2} found a theoretical
support to his previous results by constructing an equation of the
Schr\"{o}dinger type. In this way, he tried to introduce the
macroquantization of orbits in a gravitational field.\\

In the last years the idea that of a quantization of the
gravitational field has been constated. In words of Halton Arp
(see \cite{A}): {\it An unexpected property of astronomical
objects (and therefore an ignored and suppressed subject) is that
their properties are quantized, for instance the redshifts of
galaxies. The most astonishing result was then pointed to by Jess
Artem, that the same quantization ratio that appeared in quasar
redshifts appeared in the orbital parameters of the planets in the
solar system. Shortly, afterward Oliveira Neto {\it et al.}
\cite{ON}, Agnese and Festa \cite{AF}, L. Nottale {\it et al.}
\cite{N1,N2} and A. and J. Rub\v{c}i\'{c} \cite{RR1,RR2}
independently in Brazil, Italy, France and Croatia began pointing
out similarities to the Bohr atom in the orbital placement of the
planets. Different variations of the Bohr-like $r_n = n^2$ or
$r_n=n^2 + n/2$ fit the planetary semimajor axes extremely well
with rather low "quantum" numbers $n$. It is clear that the
properties of the planets are not random and that they are in some
way connected to quantum mechanical parameters both of
which are connected to cosmological properties.}\\

Action at distance in Newtonian physics is replaced by finite
propagation speeds in classical post--Newtonian physics. As a
result, the differential equations of motion in Newtonian physics
are replaced by functional differential equations, where the delay
associated with the finite propagation speed is taken into
account. Newtonian equations of motion, with post--Newtonian
corrections, are often used to approximate the functional
differential equations, see, for instance,
\cite{Ch,Ch1,Ch2,Ch3,G1,R,R2}. In \cite{G2} a simple atomic model
based on a functional differential equation which reproduces the
quantized Bohr atomic model was presented. The unique assumption
was that the electrodynamic interaction has finite propagation
speed, which is a consequence of the Relativity theory. An
straightforward consequence of the theory developed in \cite{G2},
and taking into account that gravitational interaction has also a
finite propagation speed, is that the same model is applicable to
the gravitational 2-body problem. In the following section we
present a simple gravitational model based on a functional
differential equation which gives an explanation of the modified
Titius--Bode law.

\section{The retarded gravitational 2-body problem}

We consider two particles of masses $m$ and $M$, with $m \ll M$,
interacting through the retarded inverse square force. The force
on the mass $m$ exerted by the mass $M$ is given by
\begin{equation}
{\bf F}=G \frac{M \, m}{r^3} \ {\bf r}. \label{fo}
\end{equation}

\begin{figure}[htb]
\centerline{\hbox{
\epsfig{file=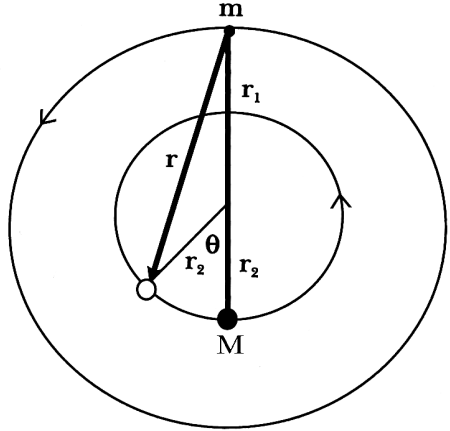,height=6.0cm,width=6.0cm} }} \caption{The
retarded gravitational 2-body problem.} \label{fig1}
\end{figure}
\par

The force acts in the direction of the 3--vector ${\bf r}$, along
which the mass $M$ is "last seen" by the mass $m$. The 3--vector
${\bf r}$ may be represented by
\[
{\bf r}= {\bf r}_M(t-\tau) - {\bf r}_m(t),
\]
where ${\bf r}_M(t)$ and ${\bf r}_m(t)$ denote respectively the
instantaneous position vectors of the mass $M$ and the mass $m$,
respectively, at time $t$, and $\tau$ is the delay, so that ${\bf
r}_M(t-\tau)$ is the "last seen" position of the mass $M$.
Assuming that the two bodies are in rigid rotation with constant
angular velocity $\omega$, and referring back to Fig. 1, we have,
in 3--vector notation,
\[
{\bf r}_m=r_1[\cos \omega t \ {\bf \hat{\i}} + \sin \omega t \
{\bf \hat{\j}}],
\]
and
\[
{\bf r}_M=-r_2[\cos \omega (t-\tau) \ {\bf \hat{\i}} + \sin \omega
(t-\tau) \ {\bf \hat{\j}}].
\]
Hence, the 3--vector ${\bf r}$ is given by
\[
{\bf r}=[-r_2\cos \omega (t-\tau) -r_1\cos \omega t] \ {\bf
\hat{\i}} + [-r_2 \sin \omega (t-\tau) -r_1 \sin \omega t] \ {\bf
\hat{\j}},
\]
Now, we introduce the polar coordinates $(r, \theta)$ and define
the unitary vectors ${\bf l}= \cos \theta \ \hat{\i} + \sin \theta
\ \hat{\j}$ and  ${\bf n}= - \sin \theta \ \hat{\i} + \cos \theta
\ \hat{\j}$. By straightforward calculations it is easy to see
that the components of the force \eqref{fo} in the polar
coordinates are
\[
F_r=G \frac{M \, m}{r^3}{\bf r} \cdot {\bf l} = (-r_2 \cos (\omega
\tau) - r_1) G \frac{M \, m}{r^3} \,
\]
and
\[
F_\theta=G \frac{M \, m}{r^3}{\bf r} \cdot {\bf n} = r_2 \sin
(\omega \tau) G \frac{M \, m}{r^3} \,
\]
The equations of the movement are
\begin{eqnarray}
m \ddot{r}- m r \dot{\theta}^2 = F_r, \label{F1} \\
m r \ddot{\theta} + 2 m \dot{r} \dot{\theta} = F_\theta.
\label{F2}
\end{eqnarray}
The second equation \eqref{F2} can be written in the form
\begin{equation}
\frac{1}{r}\frac{d L}{dt}= \frac{1}{r}\frac{d}{dt}(m r^2 \dot
\theta) = F_\theta=  r_2 \sin (\omega \tau) G \frac{M \, m}{r^3}.
\label{FF2}
\end{equation}
If we accurately study equation \eqref{FF2} we see that the
analytic function $\sin (\omega \tau)$ has a numerable number of
zeros given by
\begin{equation}
\omega \tau = k \pi \, , \label{eeqq}
\end{equation}
with $k \in \Z$, which are stationary orbits of the system of
equations \eqref{F1} and \eqref{F2}. When $\omega \tau \ne k \pi$
we have a torque which conduces the mass $m$ to the stationary
orbits without torque, that is, with $\omega \tau =k \pi$. In fact
the stationary orbits are limit cycles in the sense of the
qualitative theory developed by Poincar\'e, see \cite{P1}.

This is a new form of treating the gravitational 2-body problem
from a dynamic point of view instead of from a static point of
view, as it has been made up to now. Moreover, in this model the
delay $\tau$ is not small, in fact
\[
\tau = \frac{k \pi}{\omega}= \frac{k \pi}{2 \pi/ T} = \frac{k
T}{2} \, ,
\]
where $T$ is the time taken by the mass $m$ to complete its orbit,
i.e., the period of revolution. Therefore, the delay is a multiple
of the half--period $T/2$.

On the other hand, in a first approximation, the delay $\tau$ can
be equal to $r/c$ (the time that the field uses to goes from the
mass $M$ to the mass $m$ at the speed of the light). In this case,
from equation \eqref{eeqq} we have
\begin{equation}
\tau= \frac{k \pi}{\omega} = \frac{r}{c} \, . \label{eeqq2}
\end{equation}
Taking into account that $\omega = v_{\theta} /r$, from
\eqref{eeqq2} we have $v_{\theta}/ c = k \pi$. However, from the
Relativity theory we know that $v_{\theta}/ c < 1$, then we must
introduce a new constant $g$ in the delay. Hence, $\tau= g \, r/c$
and the new equation \eqref{eeqq2} is
\begin{equation}
\tau= \frac{k \pi}{\omega} = \frac{g \, r}{c} \, , \label{eeqq3}
\end{equation}
and now $v_{\theta}/ c = k \pi / g$, i.e. $v_{\theta}= k \pi c /
g$ and from \eqref{eeqq3} we also have $r= k \pi c /(g \omega)$.
In our model case of a classical rigid rotation we have $\theta=
\omega t$ with $\omega
> 0$. Therefore, $\dot \theta = \omega$ and $\ddot{\theta}=0$.
Hence, equation \eqref{F2} for $\omega \tau = k \pi$ is
\[
2 m \dot{r} \omega =0,
\]
which implies $\dot{r}=0$ and $r=r_k$ where $r_k$ is a constant
for each $k$. On the other hand, equation \eqref{F1} for $\omega
\tau = k \pi$ takes the form:
\begin{equation}
-m r \dot \theta^2 = -m \frac{v_{\theta}^2}{r} = (-r_2(-1)^n-r_1)
G \frac{M \, m}{r^3} \approx -r \ G \frac{M \, m}{r^3}, \label{F3}
\end{equation}
assuming that $r \sim r_1$ due to $r_2 \ll r_1$ in the case that
$m \ll M$.

From the definition of angular momentum $L=m r^2 \dot \theta= m
r^2 \omega = m r v_{\theta}$ we have that $v_{\theta}=L/(m r)$.
Substituting this value of $v_{\theta}$ into equation \eqref{F3}
we obtain $r=L^2/(G M \, m^2)$. The energy of the mass $m$
(substituting the values of $v_{\theta}$ and $r$) is given by
\begin{equation}
E=\frac{mv_{\theta}^2}{2}- G \frac{M \, m}{r}=-\frac{G^2 M^2 \,
m^3}{2 \, L^2}. \label{F5}
\end{equation}
The angular momentum for $\omega \tau = k\pi$ is
\begin{equation}
L= m v_{\theta} r = m \frac{k \pi c}{g} \frac{L^2}{G M \, m^2},
\label{F4}
\end{equation}
which is an equation for the angular momentum. Isolating the value
of $L$ we obtain $L=(G M \, m \, g)/(k \pi c)$. If we introduce
this value of the angular momentum in the expression of the energy
\eqref{F5} we have
\begin{equation}
E= -\frac{m \pi^2 c^2}{2 g^2} k^2. \label{F9}
\end{equation}
The adimensional constant $g$ cannot be unambiguously determined,
and it must be computed according with the experimental data.
However, one may speculate by considering the similarity between
the gravitational constant force $F_g$ and the electrodynamic
force $F_e$. The absolute ratio of the forces is
\[
\frac{F_g}{F_e}=\frac{G M \, m}{e^2/ 4 \pi \varepsilon_0},
\]
where $\varepsilon_0$ is the electric permittivity constant of
vacuum. By introducing the well-known fine structure constant
$\alpha$ defined by
\[
\alpha= \frac{e^2}{4 \pi \varepsilon_0 \hbar c},
\]
the ratio may be expressed by
\[
\frac{F_g}{F_e}=\frac{1}{\alpha} \left (\frac{G M \, m}{\hbar c}
\right)  =\frac{\alpha_g}{\alpha},
\]
where $\alpha_g$ is the adimensional gravitational fine structure
constant, see \cite{MTW}. However, this value of the adimensional
gravitational fine structure constant does not agree with the
experimental data given in the works \cite{AF,N1,N2,ON,RR1,RR2}.
If we recall the expression of the energy levels for the
electrostatic interaction given by Bohr in 1913
\[
E= - \frac{1}{2} \, m \, (\alpha c)^2 \frac{1}{n^2},
\]
a straightforward generalization gives us that the expression of
the energy levels for the gravitational interaction
\begin{equation}
E= - \frac{1}{2} \, m \, (\alpha_g c)^2 \frac{1}{n^2}. \label{F10}
\end{equation}
Therefore, comparing with \eqref{F9} (identifying $n=|k|$) we must
impose that the constant $g=k^2g_1$ where $g_1=\pi / \alpha_g$ and
then the energy takes the form \eqref{F10}. Therefore we have
found the value of the adimensional constant $g_1$ and
consequently the expression of the delay $\tau$ which is
\begin{equation}
\tau= \frac{g \, r}{c}= k^2 \, \frac{g_1 r}{c}= k^2 \, \frac{\pi
r}{\alpha_g c}. \label{re}
\end{equation}
In fact, we generalize the expression of the delay for the
electrodynamic interaction given in \cite{G2}. From the found
value of the angular momentum and the value of $v_{\theta}=
\alpha_g c /k$ we have
\begin{equation}
L= \frac{G \,M \, m \,  g}{k \pi c}=\frac{ G \,M \, m \,
k}{\alpha_g \, c} = m v_{\theta} r = m \frac{\alpha_g \, c}{k} r.
\label{F7}
\end{equation}
Isolating the value of $r$ from equation \eqref{F7} and
identifying $n=|k|$, we arrive to the radii of the stationary
orbits
\begin{equation}
r=\frac{G \,M \, n^2}{\alpha_g^2 \, c^2}, \label{rr}
\end{equation}
which depends on $M$. Equation \eqref{rr} agrees with the
experimental data given in the works \cite{AF,N1,N2,ON,RR1,RR2},
where $\alpha_g$ is empirically determined. One of the most
important differences between the electrodynamic interaction and
the gravitational interaction is that the unit of mass starts from
zero but the basic unit of charge never changes. For this reason
it was difficult to detect the gravitational quantization. The
dependence of the stationary orbits on $M$ and the possible
dependence of $\alpha_g$ on the masses $M$ and $m$ makes it
difficult to find a unique law for the distribution of planets in
the solar system. This is the reason that implies the using, in
some models, different expressions for the interior planets and
the exterior planets, see \cite{RR1,RR2}. If all the planets had
the same mass $m$ it would be easy to find a general law of
distribution similar to the quantization given in atomic models.

As a consequence of the values of $v_{\theta}$ and $r_n$ we have
that the period $T$ of revolution of the planets is proportional
to $n^3$ because
\[
T_n = \frac{2 \pi r}{v_\theta}= \frac{ 2 \pi \frac{G \,M \,
k^2}{\alpha_g^2 \, c^2}}{\frac{\alpha_g \, c}{k}}= \frac{2 \pi G
\, M \, k^3 }{\alpha_g^3 c^3} =  \frac{2 \pi G \, M \, n^3
}{\alpha_g^3 c^3}.
\]
Therefore, as it must happen, it is satisfied the third Kepler's
law, i.e., the ratio $r_n^3/T_n^2$ does not depend on $n$.

Summarizing, with the found delay definition \eqref{re}, the model
presented in this work explains the modified Titius--Bode law
faithfully. The gravitational quantization is, in fact, the first
approximation in the value $v/c$ of the delay in the gravitational
interaction. This first approach to the macroquantization of
orbits is confirmed by the observed data analyzed in Oliveira Neto
{\it et al.} \cite{ON}, Agnese and Festa \cite{AF}, L. Nottale
{\it et al.} \cite{N1,N2} and A. and J. Rub\v{c}i\'{c}
\cite{RR1,RR2}. In these works different values of $\alpha_g$ are
found. For instance, in \cite{RR1,RR2} the value of the
gravitational fine structure constant is $\alpha_g=\alpha / (2 \pi
f )$, where $\alpha$ is the electrodynamic fine structure
constant, and $f$ is an adimensional constant determined for each
planet. In \cite{AF} the value of the gravitational fine structure
constant determined according with the experimental data is
$\alpha_g=1/2086$, which explains the distribution of all the
planets in the solar system. An open important problem is the
determination of $\alpha_g$ in function of both masses $M$ and $m$
and other universal constants.

\section{Concluding remarks}

In \cite{G2} the atomic Bohr model is completely described by
means of functional differential equations. It is important to
stand out that what we will carry out in the following section is
not a post--Newtonian approach in which $\tau$ is small. This is
what has been made up to now and in the mentioned works
\cite{Ch,Ch1,Ch2,Ch3}. In this work, we will accept that the laws
governing the movement have a delay (a delay that does not need to
be small) and we will find a solution of the functional
differential equation in a very simple case. In this work we have
obtained the Newtonian approximation of the gravitational field
taking into account that the gravitational interaction has finite
propagation speed, which is a consequence of the Relativity
theory. This Newtonian approximation is a reminiscent of
quantization of the gravitational field. The quantization of the
gravitational field must be obtained using the Einstein's field
equation and the delay, which must appear in a natural way in this
equation.\\

\noindent{\bf Acknowledgements:}

The author would like to thank Prof. M. Grau from Universitat de
Lleida for several useful conversations and remarks.


\begin{thebibliography}{99}

\bibitem{AF} {\sc A.G. Agnese and R.Festa}, {\it Discretization on
the cosmic scale inspired from the Old Quantum Mechanics},
Hadronic J. {\bf 21} (1998), 237--253.

\bibitem{A} {\sc H. Arp}, http://www.haltonarp.com/?Page=Abstracts\&ArticleId=2

\bibitem {Ch} {\sc C. Chicone}, {\it What are the equations of
motion of classical physics?}, Can. Appl. Math. Q. {\bf 10}
(2002), no. 1, 15--32.

\bibitem {Ch1} {\sc C. Chicone, S.M. Kopeikin, B. Mashhoon and D. Retzloff}, {\it Delay
equations and radiation damping}, Phys. Letters {\bf A 285}
(2000), 17--16.

\bibitem {Ch2} {\sc C. Chicone}, {\it Inertial and slow manifolds for
delay equations with small delays}, J. Differential Equations {\bf
190} (2003), no. 2, 364--406.

\bibitem {Ch3} {\sc C. Chicone}, {\it Inertial flows, slow flows, and
combinatorial identities for delay equations}, J. Dynam.
Differential Equations {\bf 16} (2004), no. 3, 805--831.

\bibitem{G1} {\sc J. Gin\'e}, {\it \ On the classical descriptions of the
quantum phenomena in the harmonic oscillator and in a charged
particle under the coulomb force}, Chaos Solitons Fractals {\bf
26} (2005), 1259--1266.

\bibitem{G2} {\sc J. Gin\'e}, {\it \ On the origin of quantum
mechanics}, physics/0505181, preprint, Universitat de Lleida,
2005.

\bibitem {Gu1} {\sc Yu. K. Gulak}, Astrometria i Astrofizika {\bf 16} (1972), 92
(in russian).

\bibitem {Gu2} {\sc Yu. K. Gulak}, Astron. Zhurn. {\bf 57} (1980),
142 (in russian).

\bibitem{HT} {\sc W. Hayes and S. Tremaine}, {\it Fitting selected random
planetary systems to Titius--Bode laws}, Icarus {\bf 135} (1998),
549--557.

\bibitem{LP} {\sc J. Llibre and C. Pi\~{n}ol}, {\it A gravitational
approach to the Titius-Bode law}, Astronomical Journal {\bf 93}
(1987), 1272--1279.

\bibitem{MTW} {\sc C.W. Misner, K.S. Thorne and J.A. Wheeler}, {\it
Gravitation}, Freeman and Comp., San Francisco, 1973, p. 412.

\bibitem{N1} {\sc L. Nottale}, {\it Scale relativity and
quantization of extra-solar planetary systems}, Astron. Astrophys.
Lett. {\bf 315} (1996), L09--L12.

\bibitem{N2} {\sc L. Nottale, G. Schumacher  and J. Gay}, {\it Scale
relativity and quantization of the Solar System}, Astron.
Astrophys. {\bf 322} (1997), 1018--1025.

\bibitem{ON} {\sc M. de Oliveira Neto, L. A. Maia and S. Carneiro},
{\it An alternative theoretical approach to describe planetary
systems through a Schrodinger-type diffusion equation}, Chaos,
Solitons Fractals {\bf 21} (2004), 21--28.

\bibitem{P1} {\sc H. Poincar\'e}, {\it M\'emoire sur les courbes
d\'efinies par les \'equations diff\'erentielles.} Journal de
Math\'ematiques {\bf 37} (1881), 375-422; {\bf 8} (1882), 251-296;
Oeuvres de Henri Poincar\'e, vol. I, Gauthier-Villars, Paris,
(1951), pp. 3-84.

\bibitem{R} {\sc C.K. Raju}, {\it The electrodymamic 2-body problem
and the origin of quantum mechanics}, Foundations of Physics {\bf
34} (2004), 937--962.

\bibitem{R2} {\sc C.K. Raju}, {\it Time: towards a consistent theory},
Kluwer academic, Dordrecht, 1994.

\bibitem{RR1} {\sc A. Rub\v{c}i\'{c} and J.
Rub\v{c}i\'{c}}, {\it Stability of gravitational-bound many-boy
systems}, Fizika B (Zagreb) {\bf 4} (1995), 11--28.

\bibitem{RR2} {\sc A. Rub\v{c}i\'{c} and J.
Rub\v{c}i\'{c}}, {\it Square law for orbits in extra--solar
planetary systems}, Fizika A (Zagreb) {\bf 8} (1999), 45--50.

\bibitem{T} {\sc L.J. Tomley}, {\it Bode's law and the ''missing moons''
of Saturn}, Am. J. Phys {\bf 47} (1979), 396--398.

\end{thebibliography}
\end{document}